\documentclass[twocolumn,floatfix,groupedaddress,superscriptaddress,prb,showpacs]{revtex4}

\usepackage{graphicx}
\usepackage{amsmath}
\usepackage{amssymb}
\usepackage{amsfonts}
\usepackage{dcolumn}
\usepackage{bbm}
\usepackage{float}

\usepackage{color}

\begin{document}
\title{Interaction quenches in the two-dimensional fermionic Hubbard model}

\author{Simone A. Hamerla}
\email{simone.hamerla@tu-dortmund.de}
\affiliation{Lehrstuhl f\"{u}r Theoretische Physik I, 
Technische Universit\"{a}t Dortmund,
 Otto-Hahn Stra\ss{}e 4, 44221 Dortmund, Germany}

\author{G\"otz S. Uhrig}
\email{goetz.uhrig@tu-dortmund.de}
\affiliation{Lehrstuhl f\"{u}r Theoretische Physik I, 
Technische Universit\"{a}t Dortmund,
 Otto-Hahn Stra\ss{}e 4, 44221 Dortmund, Germany}

\date{\textrm{\today}}

\begin{abstract} 
The generic non-equilibrium evolution of a strongly interacting  fermionic system is studied.
For strong quenches, a collective collapse-and-revival phenomenon is found extending over 
the whole Brillouin zone. A qualitatively distinct behavior occurs for weak quenches
where only weak wiggling occurs. Surprisingly, no evidence for prethermalization is found
in the weak coupling regime. In both regimes, indications for relaxation beyond
oscillatory or power law behavior are found and used to estimate relaxation rates without resorting to
a probabilistic ansatz. The relaxation appears to be fastest for intermediate
values of the quenched interaction.
\end{abstract}


% 05.70.Ln     Nonequilibrium and irreversible thermodynamics
% 71.10.Fd     Lattice fermion models (Hubbard model, etc.)
% 67.85.-d     Ultracold gases, trapped gases
% 71.10.Pm     Fermions in reduced dimensions 

\maketitle

%%%%%%%%%%%%%%%%%%%%%%%%%%%%%%%%%%%%%
%%%%%%%%%%%%%%%%%%%%%%%%%%%%%%%%%%%%%%%%%%%%%%%%%%%%%%%%%%%%%%%%%%%%%%%%%%%%%%%
\section{Introduction}
\label{sec:intro}

Recently refined experimental techniques based on ultracold gases in optical lattices 
\cite{grein02b,kinos06} and femtosecond spectroscopy \cite{perfe06} allow for studies of systems out of
equilibrium. Such studies require a very good decoupling from the environment to realize long observation 
times during which the system is out of equilibrium. 
One way to push the system far out of equilibrium is to switch
intrinsic system parameters abruptly. {Such a scenario is called a quench}. 
In interaction quenches the system is prepared initially in 
eigenstates of a non-interacting Hamiltonian. At a specific 
time the interaction is suddenly turned on and the state of the system
is no longer an eigenstate of the (quenched) Hamiltonian. 

Typically, the quenched systems  are in highly excited states with respect
to the quenched Hamiltonian. Thus their dynamics is governed by processes 
on all energy scales including high energies. Properties 
may occur which are totally different from the equilibrium ones.
The necessity to include all energy scales makes theoretical calculations, numerical or analytical ones,
notoriously difficult. So far, the majority of  theoretical investigations were
focussed on one-dimensional (1D) systems, on infinite-dimensional ($\infty$D) systems,
and on small finite systems because for these cases powerful tools
are available. For 1D systems, the tool box is best: Quantum field theoretical
descriptions provide analytical approaches, see, e.g., Refs.\ 
\onlinecite{cazal06,uhrig09c,fiore10,sabio10,schur12,rentr12}. The best understood models
remain those which correspond to non-interacting fermionic or bosonic systems
\cite{barth08,iucci10,calab12a,calab12b} or models which
are effectively close to non-interacting ones \cite{kenne13}.
Time-dependent density-matrix renormalization
is a powerful numerical tool which enables to study non-equilibrium phenomena
in 1D systems \cite{daley04,white04a,manma07,karra12b}.
The other dimensionality allowing for well-controlled studies 
is $\infty$D where dynamical mean-field theory becomes exact
\cite{schmi02,freer06,eckst09} and Gutzwiller approaches are well justified \cite{schir10}.
Exact diagonalization is completely flexible concerning 
dimensionality, but it is restricted to small systems \cite{rigol08,torre13}.

So far, the question to which extent strongly conserved quantities restrict or even
prevent relaxation was in the center of interest
\cite{rigol07,fiore10,kolla11,polko11,calab12a,calab12b}.
Thus, integrable systems and systems close to integrability were studied, which drew the interest
to 1D systems.
Studies of two-dimensional (2D) models out of equilibrium are still rare. Goth and Assaad studied the 
sudden turning off of the interaction in a half-filled 2D Hubbard model with 20$\times$20 sites by 
continuous time quantum Monte Carlo \cite{goth12}. The large energy put into the
system and the simple dynamics induced by the non-interacting Hamiltonian after the quench
lead to a well-understood evolution in agreement with the findings of perturbative approaches.
Other 2D studies address the influence of a strong electric field on the dynamics of a single
charge carrier in a Mott insulator \cite{mierz11} and on a bound pair of two carriers \cite{bonca12}.

In the present work, we study the interaction quench
in the 2D Hubbard model far from any integrability.
{In contrast to the work by Goth and Assaad, the interaction is switched \emph{on} abruptly.}
Our goal is to assess the time scale on which relaxation takes place in a generic 
model between one and infinite dimension. The sensitive quantity which we investigate
is the momentum distribution $n_k(t):=\langle c^\dagger_{k,\sigma} c_{k,\sigma}\rangle$
and its jump $\Delta n_k(t)$ at the Fermi surface $k=k_\text{F}$ in particular.

The article is set up as follows. After this introduction we {present} the model studied
and the method used in Sect.\ \ref{sec:mod_n_meth}. In the subsequent Sect.\ \ref{sec:results}
the results for the momentum distribution and its jump 
at the Fermi surface will be presented. The scenario for prethermalization
will be an important issue as well as an estimate of relaxation rates.
In Sect.\ \ref{sec:conclude} the article is concluded.

\section{Model and Method}
\label{sec:mod_n_meth}

The Hamiltonian under study is 
\begin{align}
\hat{H} = -J \sum_{\langle\vec{r},\vec{s}\rangle,\sigma}
(\hat{c}_{\vec{r},\sigma}^\dagger\hat{c}_{\vec{s},\sigma}^{\phantom\dagger}+\text{h.c.}) 
+ U(t) \sum_{\vec{r}}:\hat{n}_{\vec{r},\uparrow}\hat{n}_{\vec{r}\downarrow}:
\end{align}
with the hopping parameter $J$; $\vec{r}$ and $\vec{s}$ denote nearest neighbors on the square lattice. The fermionic annihilation (creation) operator at site $\vec{r}$ with spin $\sigma$ is denoted by $\hat{c}_{\vec{r},\sigma}^{(\dagger)}$ and $\hat{n}_{\vec{r},\sigma}$ counts the fermions at site $\vec{r}$. 
Profiting from  translational invariance we directly address the infinite model in the thermodynamic limit. The colons indicate normal ordering with respect to the Fermi sea, which is the
ground state of the {non-interacting} model and the initial state of the quench.
The interaction $U(t) = U\Theta(t)\geq 0$ is suddenly turned on at $t=0$ so that
the time evolution is governed by the interacting Hamiltonian. 

We use the band width $W=8J$ as a natural energy scale and $\hbar$ is set to unity 
so that time is measured in the inverse band width $1/W$. The time evolution of the jump $\Delta n_k(t)$ is used as sensitive probe for the dynamics after the quench. Its initial value is unity.
The momentum distribution 
is calculated by an expansion of the Heisenberg equations of motion (EoM) for an operator $\hat{A}$ 
\begin{align}
\partial_t \hat{A}(\vec{r},t) = i\big[\hat{H},\hat{A}(\vec{r},t)\big]
\end{align}
to the highest order possible \cite{uhrig09c}. We consider $\hat A(t=0)=c^\dagger_{k,\sigma}$.
The EoM are iterated by recursive commutation with $H$ 
yielding more and more operators which we represent in real space \cite{hamer13a, hamer13b}. 
As each commutation implies one additional order in time $t$ the results obtained after $n$ commutations are  exact at least up to $t^n$. 

We emphasize, however, that we are not computing a plain series in powers of time.
In the algebraic part, we derive a set of differential equations which allow for
the determination of the series up to $t^n$. This means that the solution of the
approximate differential equations has the same expansion in powers of $t$ as the exact
solution. But here we do not present results for the series, but for the full solution
of the approximate differential equation which turn out to be more stable and reliable
up to longer times than the plain series. Thus the order $n$ of the calculation
is a control parameter of the approximation which becomes exact for $n\to\infty$,
but it does not refer to the maximum order of a truncated series.

Due to the exponentially rising number of terms to be tracked for increasing $n$
one has to stop at values of $n$ of the order 10.
The results are well-controlled \cite{hamer13a, hamer13b} for about $t\lessapprox n/W$ .
 For the 2D model, up to $n= 9$ commutations are performed and the data is shown 
 up to times for which the results are reliable. This can be inferred from the 
 comparison of the curves for various numbers $n$, which display convergence upon
 increasing $n$, see also Refs.\ \onlinecite{hamer13a,hamer13b}.

\section{Results}
\label{sec:results}

The approach sketched above is applied to the fermionic creation operator.
In this way, the time dependence of expectation values such
as $\langle c_{\vec{r}} c_{\vec{s}}^\dag\rangle(t)$ becomes accessible.
Fourier transformation of these expressions yields the momentum distribution.

\subsection{Momentum Distribution}
\label{ssc:md}

\begin{figure}[htb]
    \begin{center}
    \includegraphics[width=1.0\columnwidth,clip]{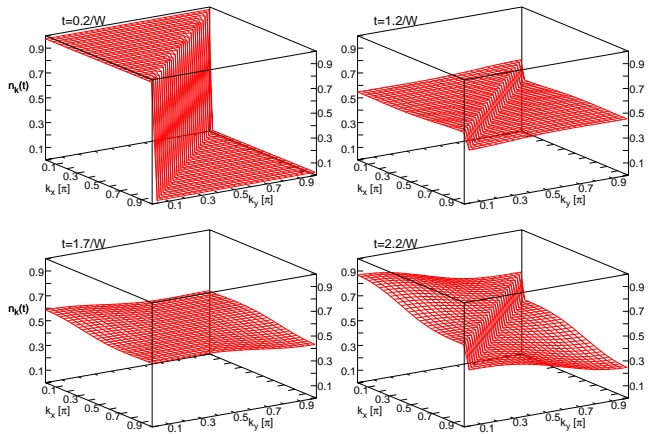}
    \end{center}
    \caption{(Color online) Evolution of the momentum distribution of the half-filled
    Hubbard model quenched to $U=2W$. Only one quadrant of the Brillouin zone is shown
    due to point group symmetry. The data results from $n=6$ commutations.
    \label{fig:md}
 }
\end{figure}

We show a complete view on the momentum distribution in the Brillouin zone
 in Fig.\ \ref{fig:md} as function of time. Since many points in momentum space have
 to be evaluated we have to restrict ourselves to $n=6$ commutations. Thus the 
 data {in Fig.\ \ref{fig:md}} is not of the highest accuracy for the longer time,
 but it renders an excellent overview. {All other figures present data for $n=9$ commutations.}
 
 For the relatively large values $U=2W$, the momentum distribution {in Fig.\ \ref{fig:md}} 
 shows oscillations over the whole Brillouin zone. At the instants
 at which the jump vanishes, see panel $t=1.7/W$, the distribution is featureless and almost
 constant indicating a state which is essentially local in real space. 
 But afterwards, the jump re-occurs and the momentum distribution
 resembles the initial one qualitatively. Thus the total behavior follows a 
 collapse-and-revival scenario.  
 
 The results in Fig.\ \ref{fig:md} suggest
 that it would be very fascinating to observe such a behavior in fermionic systems experimentally.
 Note that collapse-and-revival was observed experimentally after the interaction quench
 in a bosonic system \cite{grein02b}. But there is an essential qualitative difference 
 between the fermionic and the bosonic collapse-and-revival. As seen in 
 Fig.\ \ref{fig:md}, the fermionic one is characterized by the disappearance and
 re-appearance of the Fermi surface, i.e., a one-dimensional singularity in the 
 two-dimensional Brillouin zone.
 In contrast, the bosonic collapse-and-revival is related to the disappearance
 and re-appearance of a zero-dimensional singularity, namely of a
 $\delta$-function in the bosonic momentum distribution at the center ($\Gamma$ point)
 of the Brillouin zone.
 
\begin{figure}[htb]
    \begin{center}
    \includegraphics[width=1.0\columnwidth,clip]{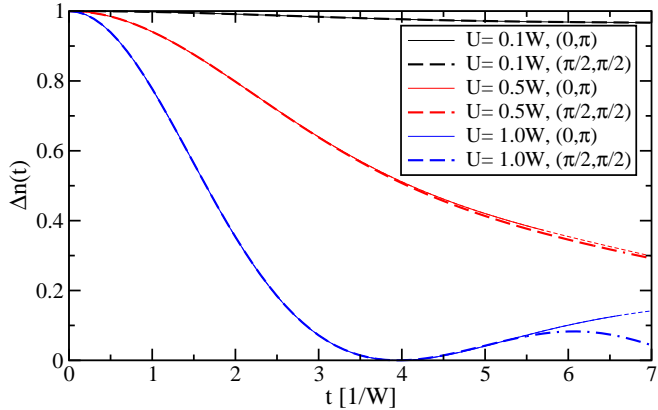}
    \end{center}
    \caption{(Color online) Jump $\Delta n(t)$ calculated at two positions
    on the Fermi surface for various interaction strengths $U$. The curves for $U=1.0W$
    change their style for larger times; then we do not consider them fully
    reliable anymore.
    \label{fig:momentum-depend}
 }
\end{figure}

 We find that the jump $\Delta n_{k_\text{F}}(t)$ behaves very similar at
 all points of the Fermi surface. No significant difference between the 
 jumps at the corners of the Fermi surface, i.e., at $k=(\pm\pi,0)$ and $(0,\pm\pi)$, 
 and those at the middle of the edges, i.e., at $k=(\pm \pi/2,\pm \pi/2)$.
 appears up to the time-scales investigated. This is illustrated in Fig.\ 
 \ref{fig:momentum-depend} for the jump $\Delta n(t)$ at the given momenta
 on the Fermi surface, see legend.  
 Note that the difference between curves for different momenta first increases 
 on increasing $U$ before decreasing again for larger $U$. In any case, it remains
 small even for $U=0.5W$ up to the time scales investigated.
 In the remainder, we will only show results for $k=(\pi,0)$ for simplicity
 if not stated otherwise.

\subsection{Comparison to the behavior in 1D}
\label{ssc:vgl-1D}

\begin{figure}[htb]
    \begin{center}
    \includegraphics[width=1.0\columnwidth,clip]{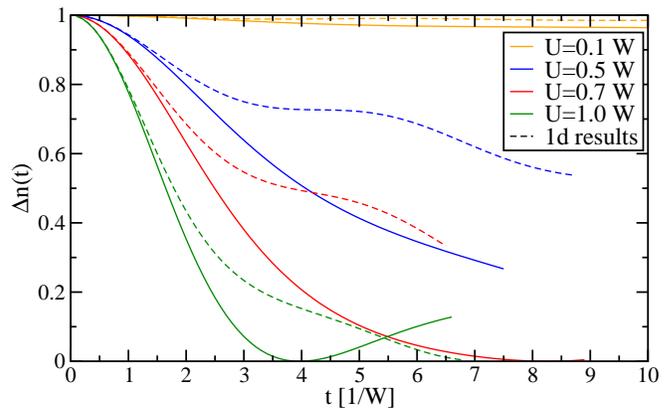}
    \end{center}
    \caption{(Color online) Comparison of the time dependence of the jump $\Delta n(t)$ 
    for various $U$ in half-filled Hubbard models:   Solid lines show the 2D, the dashed lines of
    the corresponding gray scale/color show the 1D data. 
    \label{fig:comp2d1d}
 }
\end{figure}

In Fig.\ \ref{fig:comp2d1d} we compare the quench dynamics in 1D and in 2D 
at half-filling for various values of $U$. Our findings provide evidence that
in 2D the same dynamical transition exists between quenches to weak and to strong
interactions that was observed previously
in $\infty$D \cite{eckst09}, by Gutzwiller approach \cite{schir10}, and in 1D 
\cite{hamer13a}.
For quenches to stronger interactions ($U\gtrapprox 0.7W$) 
one observes dominant oscillations which decay slowly.
At half-filling, these oscillations display zeros in the jump $\Delta n$ as in the previous cases
\cite{eckst09,schir10,hamer13a}.
Away from half-filling, the minima still exist, but they are no longer at $\Delta n=0$ (not shown).

For quenches to weak and moderate interactions we observe a decay of $\Delta n$ 
with only hardly visible oscillations, cf.\ the curves for $U < 0.7 W$.
These oscillations can be attributed to the finite band width $W$, i.e., the frequency of oscillations
is the band width $W$. This explanation is supported by the 
fact that the oscillations are stronger in 1D than in 2D because the Van Hove
singularities in 1D (inverse square roots, $\propto \Delta\omega^{-1/2}$) \cite{rentr12,karra12b,hamer13a} 
are much more pronounced than in 2D (jumps, $\propto \Delta\omega^{0}$).
This argument is in line with the observations that no oscillations
are observed in the infinite dimensional calculations based on the Bethe
lattice with infinite branching ratio displaying even less pronounced
singularities (square roots, $\propto \Delta\omega^{1/2}$) \cite{eckst09}.

Another remarkable contrast to the 1D curves consists in the much faster decay
of the jump in 2D. This feature is striking in the curves in Fig.\  \ref{fig:comp2d1d} 
which all start with the same curvature $-U^2/2$ determined by $U$ alone. We interpret this
important qualitative difference by the fact that the decay of the jump
in 1D is governed by slowly decreasing power laws \cite{cazal06,uhrig09c,karra12b,hamer13b}.
The 2D characteristics appears different: The 2D
system allows for sufficiently effective scattering mechanisms so that one 
may expect to observe first signs of true relaxation 
governed by exponential decay  $\Delta n(t) \propto \exp(-at)$ with
relaxation rate $a>0$ for longer times. We will come back to this point below.
For intermediate values of $U\approx 0.7W$ we find a particularly fast
decaying jump indicating efficient relaxation indeed, cf.\ Fig.\ \ref{fig:comp2d1d}.

\subsection{Strong Quenches and their decay}
\label{ssc:strong}

\begin{figure}[htb]
    \begin{center}
    \includegraphics[width=1.0\columnwidth,clip]{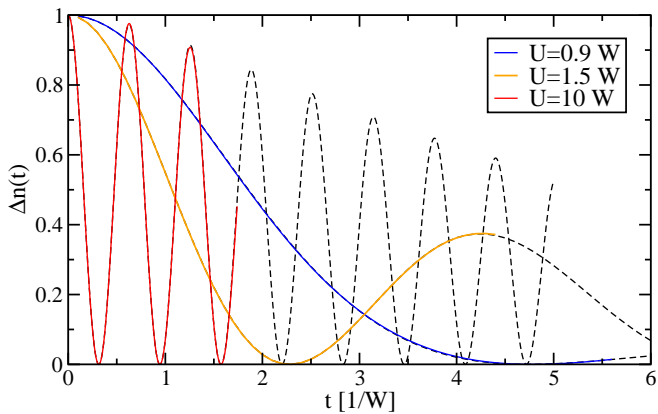}
    \end{center}
    \caption{(Color online) Fits (dashed lines) to the jump $\Delta n(t)$ from the EoM (solid lines)
    for quenches to strong interactions, fits based on Eq.\ \eqref{eq:strong}. 
      \label{fig:fits_strong}
 }
\end{figure}

\begin{figure}[htb]
    \begin{center}
    \includegraphics[width=1.0\columnwidth,clip]{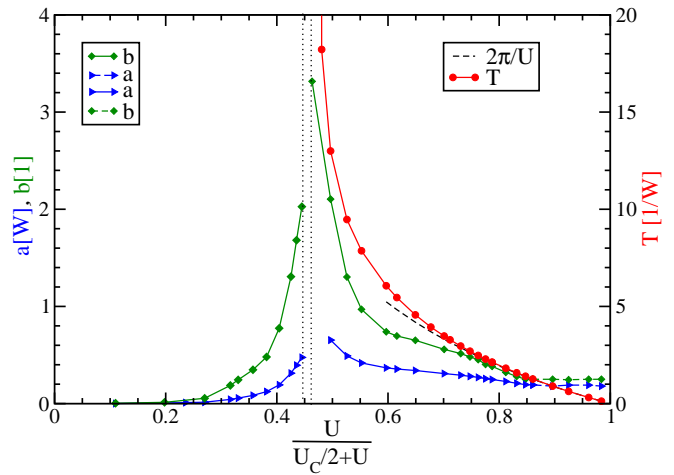}
    \end{center}
    \caption{(Color online) Fit parameters as they result from fitting the EoM data by Eq.\
   \eqref{eq:strong} for strong interactions and by  \eqref{eq:weak} for weak interactions.
   The dashed lines for $U/(U_C/2+U) \gtrapprox 0.86$ depict fits which are unstable
   and cannot be done without restrictions in the range of  parameters.
    The left scale applies to $a$ and $b$ while the right one to $T$. 
    In the transition region between
    the two dashed perpendicular lines our data does not allow us 
    to decide on the nature of the quench,
    weak or strong. The constant $U_C$ is set to $-8E_\text{kin}=(16/\pi^2)W=1.62W$.
    \label{fig:param}
 }
\end{figure}

We proceed to a quantitative analysis by fits. For strong quenches at half-filling we take the oscillations into account and we allow for relaxation to occur.
For the oscillations we simply include a cosine term, see Eq.\ \eqref{eq:strong}.
The relaxation is trickier for the following reason. In the long time limit
it is described by the factor $\exp(-at)$ with decay rate $a>0$. But around $t=0$
this behavior does not and cannot appear because the time dependence induced by Hamiltonians,
whose local terms are bounded, is analytically smooth. Thus the fit
function to describe relaxation must be smooth at $t=0$ and then it must crossover to
$\exp(-at)$. The simplest function we could think of with the desired
 property is $\exp(-\sqrt{(at)^2+b^2}+b)$. Obviously, the decay rate at large $t$ is
 given by $a$ while the behavior at small $t$ is smooth. The crossover from
 $t^2$ behavior to $|t|$ behavior occurs at  $t_\text{crossover}\approx b/a$.
Thus we use
\begin{equation}
\label{eq:strong}
\Delta n_\text{stro}(t)=\cos(\omega t)^2\exp(-\sqrt{(at)^2+b^2}+b).
\end{equation}
Two of the three fit parameters $\omega$, $a$, and $b$ are fitted and the third one is
determined by the analytic curvature $-U^2/2$ at $t=0$. Exemplary fits are shown 
in Fig.\ \ref{fig:fits_strong}. The resulting 
oscillation periods $T=2\pi/\omega$ and values for $a$ and $b$ are depicted 
 in Fig.\ \ref{fig:param} to the right of the vertical dotted lines
 which mark the region between weak and strong quenches. This region
 cannot be resolved by our present approach; the weak quenches are considered
 below in Sect.\ \ref{ssc:weak}.
 
Note that the quantitative description of our data with the fitting
function  \eqref{eq:strong} works very nicely. We are aware, however, 
that the parameters $a$ and $b$ resulting from these fits give only
estimates for the relaxation rate and the crossover time, respectively.
In case another dynamical time scale would govern the behavior at
intermediate times it may be that the numbers for $a$ and $b$ are
affected by this intermediate time scale and not by the relaxation at
long times.

\subsection{Existence of Prethermalization}
\label{ssc:pretherm}

The natural next issue are weak quenches and their decay. Indeed, we will
address it in the following subsection. But before doing so it is worth
to recall the leading perturbative result in order $U^2$ derived
by Moeckel and Kehrein \cite{moeck08,moeck09a} which reads 
\begin{equation}
\label{eq:perturb}
\Delta n_{k_\text{F},\text{2nd}}(t) =1-U^2 f_{k_\text{F}}(t)+{\cal O}(U^4)
\end{equation}
for the jump at the Fermi surface with
\begin{eqnarray}
\nonumber
&& f_{k_\text{F}}(t) = 
\\
 &&\quad \frac{4}{N^2}\sum_{p p' q} \delta_{p+k_\text{F}}^{p'+q} 
\frac{\sin^2(\Delta \varepsilon t/2)}{\Delta \varepsilon^2} (n_p \bar n_{p'}\bar n_{q}+ \bar n_p n_{p'}n_{q}), \qquad
\label{eq:multdim}
\end{eqnarray}
where $N$ is the number of sites, 
$k_\text{F}$ a wave vector on the Fermi surface, $n_p=1$ if $p$ is within the Fermi surface
and zero otherwise, $\bar n_p=1-n_p$, and $\Delta\varepsilon=\varepsilon_{k_\text{F}}+\varepsilon_p-
\varepsilon_{p'}-\varepsilon_q$. 

In infinite dimensions, the multidimensional integral \eqref{eq:multdim} is shown
to yield a constant for $t\to\infty$ so that for weak enough interaction an almost
constant plateau appears  before relaxation sets in on much longer time
scales \cite{moeck08,moeck09a,eckst09}. It is expected that this is the generic
behavior in finite dimensions as well. But it is also known from bosonization
\cite{cazal06,uhrig09c,fiore10,sabio10,schur12,rentr12}
that the 1D case is special because of particularly strong scattering yielding
a logarithmic divergence of $f_{k_\text{F}}(t)$. So the question arises
what happens in two dimensions?

\begin{figure}[htb]
    \begin{center}
    \includegraphics[width=1.0\columnwidth,clip]{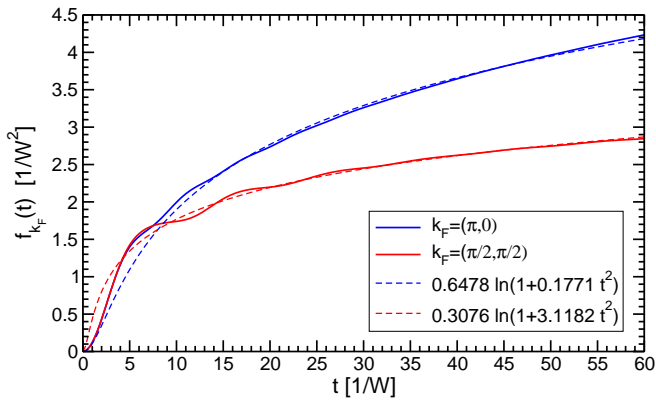}
    \end{center}
    \caption{(Color online) Leading perturbative correction
    $f_{k_\text{F}}(t)$ as it appears in \eqref{eq:perturb} and is
    defined in \eqref{eq:multdim} evaluated in 2D for the half-filled Hubbard
    model at two different momenta on the Fermi surface. Logarithmic fits
    are included.
    \label{fig:pretherm}
 }
\end{figure}

To clarify this issue we evaluated \eqref{eq:multdim} in 2D for the half-filled 
Hubbard model. This calculation is done in real space up to long times though
the limit of infinite time cannot be addressed directly. This is left to
future work. Fig.\ \ref{fig:pretherm} displays the results for 
$f_{k_\text{F}}(t)$ at momenta $(\pi,0)$ and $(\pi/2,\pi/2)$.
Unexpectedly, the data indicates a logarithmic divergence for $t\to\infty$
as is revealed by the fits. {This} suggests that no
prethermalization plateaus arise because the perturbative correction diverges
for $t\to\infty$ even for arbitrarily small quenched interaction  $U$.
Of course, this fact influences the quench dynamics decisively.

So far, we cannot prove what the reason is for the non-existence of prethermalization
in the 2D Hubbard model at half-filling. But we attribute this non-existence to the perfectly 
flat stretches of the Fermi surface linking the four points
$(\pm\pi,0)$ and $(0,\pm\pi)$. In the vicinity of these flat regions
only the perpendicular momentum transfer matters while the parallel one
can be integrated over yielding just a certain prefactor.
As a result the relevant, perpendicular scattering processes behave as
if they were acting in 1D. If this hypothesis turns out to be true,
any system doped away from half-filling should show prethermalization
because the Fermi surface will be curved.
But the time scales, on which the effect of doping becomes visible, are
presumably very long for low doping. Thus the interaction
values $U$ at which plateaus become discernible will be fairly low.
Further work is called for to elucidate this issue.

\subsection{Weak quenches and their decay}
\label{ssc:weak}

\begin{figure}[htb]
    \begin{center}
    \includegraphics[width=1.0\columnwidth,clip]{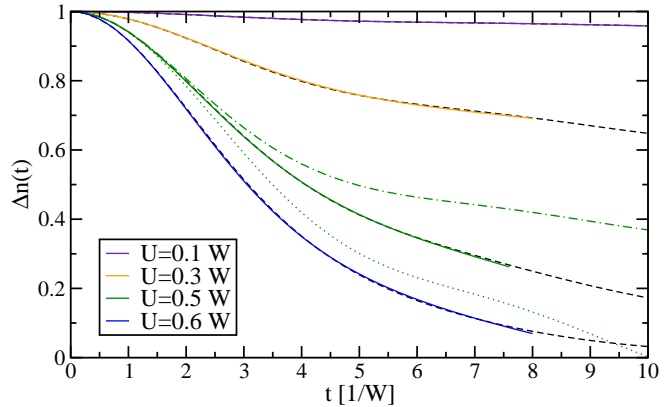}
    \end{center}
    \caption{(Color online) Fits (dashed lines) to the jump $\Delta n(t)$ from the EoM (solid lines)
    for quenches to weak interactions, fits based on Eq.\ \eqref{eq:weak}. 
    Exemplarily for $U=0.5W$, the dotted line shows 
    $\Delta n_{k_\text{F},\text{2nd}}(t)$ from Eq.\ \eqref{eq:perturb} and the 
    dashed-dotted line 
      $\Delta n_{k_\text{F},\text{exp}}(t)$ from Eq.\ \eqref{eq:exponen}.
          \label{fig:fits_weak}
 }
\end{figure}

We have argued that due the divergence of $f_{k_\text{F}}(t)$ for $t\to\infty$ the jump in \eqref{eq:perturb} does not display a prethermalization plateau \cite{moeck08,kolla11}.
A second corollary is that the strict perturbative result becomes unphysical, namely negative, see, e.g., dotted curve in  Fig.\ \ref{fig:fits_weak}. This happens
even for arbitrarily small $U$ if $t$ is chosen sufficiently large. 
We have to reconcile this behavior with the physical fact $\Delta n_{k_\text{F}}\ge 0$
because otherwise there is no way to estimate the relaxation.
Based on the analogy to the 1D case{\cite{cazal06,uhrig09c,karra12b,hamer13b}}
we propose the hypothesis that the logarithmic divergence is the signature
of a power law behavior \emph{if} there were no relaxation. 
In other words, only the deviation from the power law behavior can
be taken as sign of relaxation. 

In order to use this hypothesis we pass from the logarithmically diverging 
\eqref{eq:perturb} to  the power law behavior
\begin{equation}
\label{eq:exponen}
\Delta n_{k_\text{F},\text{exp}}(t) = \exp(-U^2f_{k_\text{F}}(t)) + {\cal O}(U^4),
\end{equation}
where we omit the corrections ${\cal O}(U^4)$.
This result only uses the leading perturbative result, but 
extrapolates it as a power law. Indeed, a comparison
to a diagrammatic analysis based on dynamic cluster theory \cite{tsuji14}
shows that weak quenches follow the prediction $\exp(-U^2f_{k_\text{F}}(t))$.
For illustration, the dashed-dotted 
curve  in Fig.\ \ref{fig:fits_weak} shows the result from \eqref{eq:exponen}
for the case $U=0.5W$.
Note that the solid curves display the full result which can be taken
to be exact up to the times shown.

The perturbative result $\Delta n_{k_\text{F},\text{exp}}(t)$ in Eq.\ \eqref{eq:exponen} serves as our reference in the following ansatz
\begin{equation}
\label{eq:weak}
\Delta n_\text{weak}(t)= \Delta n_{k_\text{F},\text{exp}}(t) \exp(-[(at)^4+b^4]^{-1/4}+b).
\end{equation}
The last factor is again chosen such that it is compatible with all known
properties of $\Delta n(t)$. It starts smoothly, all quadratic dependence in $U$ is
contained in the first factor $\Delta n_{k_\text{F},\text{exp}}(t)$ in \emph{all}
orders in $t$, and the quadratic behavior in $t$ is also fully described by 
$\Delta n_{k_\text{F},\text{exp}}(t)$ because the $U^2$ term is sufficient to 
describe the short-time behavior. This can be concluded from the EoM
used here and it was previously concluded based on other techniques \cite{moeck08}.
Thus, the minimal relaxation factor $\exp(-[(at)^4+b^4]^{1/4}+b)$ is chosen such that
it does not alter the exactly known $t^2$ term. Together with the fact
that only even powers in time and in $U$ can occur leads to the {use} of the unusual exponent
of $1/4$.

Based on \eqref{eq:weak} we fit the EoM data and determine 
$a$ and $b$ in this way. The results are shown in Fig.\ \ref{fig:param} for smaller
values of $U$, i.e., on the left side of the dashed vertical lines. 
We are aware that the decay rate $a$ and the crossover
time $b/a$ ensuing from this analysis are only estimates in view of
the  hypothesis necessary to analyze the data.
The decay rate $a$ increases only weakly for increasing $U$; our
data is consistent with  $a\propto U^4$ as it is built-in into the
fit function \eqref{eq:weak}. The $U^2$ dependence is taken into
account by the factor $\Delta n_{k_\text{F},\text{exp}}(t)$.

Our analysis is not unbiased, but relies on certain assumptions. We 
emphasize that this is also the case in many other approaches on relaxation
which rely on a probabilistic description which has relaxation built-in by construction, see
for instance \cite{erdos04,breue06}.

Except for a fairly narrow window between $U\approx0.65W$ and $0.7W$ the EoM data allows us to decide whether
a strong quench with oscillatory behavior (Eq.\ \eqref{eq:strong}) occurs or whether a 
weak quench displaying only some shoulders or wiggles occurs (Eq.\ \eqref{eq:weak}). 
The existence of these
two qualitative different regimes, separated by a dynamic transition is  obvious.
This 2D result is in line with previous observations in $\infty$D \cite{eckst09}, 
in Gutzwiller approximation \cite{schir10}, and in 1D \cite{hamer13a}.

The relaxation as captured by $a$, see Fig.\ \ref{fig:param}, is by far largest in the 
vicinity of the dynamic transition, i.e., around $U=0.7W$. In this region, the rate is of 
the order of the band width. But away from this
region, i.e., for small or for large interaction the relaxation is very weak. For low 
$U$ this is expected as explained above since the leading order $U^2$ does not lead to 
relaxation so that only the next-leading $U^4$ processes induce relaxation.

For strong values of $U$ it is remarkable that the relaxation becomes small again.
We attribute this fascinating behavior to the dominance of the local Rabi oscillations 
\cite{hamer13a} with
$\omega \approx U$, see dashed line in Fig.\ \ref{fig:param}. These oscillations do not relax
at all for $W=0$ so that the conclusion $a\propto W \propto U^0$ suggests itself.
The fits shown in Fig.\ \ref{fig:param} are consistent with this argument,
but they are not particularly stable.

\section{Conclusions and Outlook}
\label{sec:conclude}

Concluding, we studied interaction quenches in the 2D Hubbard model as a generic
finite dimensional model between one and infinite dimension. 
The momentum distribution is computed and for strong interactions
collapse-and-revival oscillations of the singular jump in the momentum distribution
at the Fermi surface is found. Though qualitatively similar to what
has been measured in bosonic systems, the key difference is that the 
bosonic collapse-and-revival occurs at a single point, the center, in the Brillouin zone.
For weak interactions, only very weak oscillations occur so that two qualitatively distinct
regimes are found, separated by a dynamic transition.

Considering the Fermi jump in the momentum distribution as particularly sensitive probe 
we have found that it decreases much faster in two dimensions than in one dimension. 
This provides evidence for relaxation without the bias of a probabilistic ansatz in terms of a density matrix. In particular for intermediate interactions $U\approx0.7W$ a significant relaxation rate
of the order of the band width was found.
Based on plausible, though not rigorous assumptions, on the functional form
of the relaxation we estimated the decay rates in the two-dimensional Hubbard model
at half-filling. The results are summarized in Fig.\ \ref{fig:param}.

By a nonequilibrium extension of dynamic cluster theory Tsuji and co-workers
have also obtained results for quenches in the two-dimensional Hubbard model \cite{tsuji14}.
Since they are using iterated perturbation theory they focus on weak quenches ($U\approx 0.25W$).
One of their central issues is the differing relaxation rate at different
points on the Fermi surface. A quantitative comparison to our approach 
shows that their data is close to what we obtain within the
exponentiated perturbative result in Eq.\ \eqref{eq:exponen}. 
But the differences for times up to $t\approx 6/W$ between
their results and the reliable results of the equation of
motion are larger than the differences between the results
for $(\pi,0)$ and $(\pi/2,\pi/2)$. Thus we consider it difficult to draw
definite conclusions on this issue at the present stage.

The approach as it is presented here 
makes contributions to nonequilibrium dynamics up to intermediate times.
Moreover, it can contribute to the theoretically fascinating, but intricate, issue 
what happens at long times by gauging other techniques. In particular, 
no assumption has been made that the evolution of the system can be described 
by a statistical mixture even though one starts from a pure state.

Furthermore,  we stress that the approach as it stands has the potential
to provide experimentally relevant data. Often, experimental
data is also restricted to short and intermediate times due to 
various disturbing effects whose detrimental influence grows in relative importance
with time.

The quenches studied in the present article started from a non-interacting
Fermi sea as an initial state which can
be treated according to Wick's theorem. 
But it must be emphasized that this property is not essential for 
the approach used. {The indispensable prerequisite is} to know the correlations
of the initial state in order that the equation of motion technique
can be put to use. Thus, many {different} initial states can indeed be treated.
Also mixtures, for instance the thermal density operator at a certain temperature
$T>0$, can be used to analyse the final result of the equations of motion.

Beyond short and intermediate times, the employed technique can be
iterated over many short time intervals to reach long times.
The key idea is to assume that a probabilistic description holds
after each short time interval so that one can re-initialize
the EoM approach after each time step.
By comparison to the direct results by EoM, one can investigate to which extent the assumption that
the system is describable as a mixture holds. 
If satisfying agreement is found one can then use the approach of iterated
time steps to reach much longer times.

Even the properties of stationary states can be tackled, that is, the steady-state
that describes the system after infinite long time. This steady-state
 can be addressed by equations of
motion if they are combined with the concept of stationary phases,
see for instance Ref.\ \onlinecite{fiore10}. In this way, 
the way is paved for the further methodological developments 
which help us to better understand nonequilibrium physics.

\begin{acknowledgments}
We are grateful for useful discussions with M.\ Eckstein and M.\ Kollar.
We acknowledge support by the Studienstiftung des deutschen
Volkes (SAH) and by the Mercator Stiftung (GSU).
\end{acknowledgments}

% \bibliographystyle{apsrev}
% \bibliography{../../bibinput/liter10} 

\end{document}